\documentclass[12pt]{article}
\begin{document}
\title{Mixed-Mode Shell-Model Calculations}
\author{V. G. Gueorguiev and J. P. Draayer \\ \\
\it Department of Physics and  Astronomy, \\
\it Louisiana State University, \\
\it Baton Rouge, LA 70803}
\maketitle

\begin{abstract}
A one-dimensional harmonic oscillator in a box is used to introduce the
oblique-basis concept. The method is extended to the nuclear shell model
by combining traditional spherical states, which yield a diagonal
representation of the usual single-particle interaction, with collective
configurations that track deformation. An application to $^{24}$Mg, using
the realistic two-body interaction of Wildenthal, is used to explore the
validity of this mixed-mode shell-model scheme. Specifically, the correct
binding energy (within 2\% of the full-space result) as well as
low-energy configurations that have greater than 90\% overlap with
full-space results are obtained in a space that spans less than 10\% of
the full-space. The theory is also applied to lower pf-shell nuclei,
$^{44-48}Ti$ and $^{48}Cr$, using the Kuo-Brown-3 interaction. These
nuclei show strong SU(3) symmetry breaking due mainly to the
single-particle spin-orbit splitting. Nevertheless, the results also show
that yrast band B(E2) values are insensitive to fragmentation of the SU(3)
symmetry. Specifically, the quadrupole collectivity as measured by B(E2)
strengths remains high even though the SU(3) symmetry is rather badly
broken. The IBM and broken-pair models are considered as alternative basis
sets for future oblique-basis shell-model calculations.
\end{abstract}
\maketitle

Typically, two competing modes characterize the structure of a nuclear system.
The spherical shell model is the theory of choice when single-particle behavior
dominates. When deformation dominates, it is the Elliott SU(3) model. This
manifests itself in two dominant elements in the nuclear Hamiltonian: the
single-particle field, $H_{0}=\sum_{i}\varepsilon _{i}n_{i}$, and a collective
quadrupole-quadrupole interaction, $H_{QQ}=Q\cdot Q$. It follows that a
simplified Hamiltonian $H=\sum_{i}\varepsilon _{i}n_{i}-\chi Q\cdot Q$ has two
solvable limits associated with these modes.

To probe the nature of such a system, we first consider the simpler problem of
a  one-dimensional harmonic oscillator in a box of size $2L$. As for real
nuclei, this system has a finite volume and a restoring harmonic
potential $\omega ^{2}x^{2}/2$. Depending on the value of $E_{c}=\omega
^{2}L^{2}/2$, which plays the role of a critical energy, there are three
spectral types: (1) for $\omega \rightarrow 0$ the spectrum is simply that of a
particle in a box; (2) at some value of $\omega$, the spectrum begins with
$E_{c}$ followed by the spectrum of a particle in a box perturbed by the
harmonic potential; and (3), for sufficiently large $\omega$ there is a
harmonic oscillator spectrum below $E_{c}$ followed by the perturbed spectrum
of a particle in a box. The last scenario is the most interesting one since it
provides an example of a two-mode system. For this case, the use of two sets of
basis vectors, one representing each of the two limits, has physical appeal.
One basis set consists of the harmonic oscillator states; the other set
consists of basis states of a particle in a box. Even thought a mixed spectrum
is expected around $E_{c}$, a numerical study that includes up to 50 harmonic
oscillator states below $E_{c}$ shows that first order perturbation theory
works well after the breakdown of the harmonic spectrum. Although the spectrum
seems to be well described in this way, the wave functions near $E_{c}$ have an
interesting coherent structure.

An application of the theory to $^{24}$Mg, using the realistic two-body
interaction of Wildenthal, demonstrates the validity of the mixed-mode
shell-model scheme. In this case the oblique-basis consists of the traditional
spherical states, which yield a diagonal representation of the single-particle
interaction, together with collective SU(3) configurations, which yield a
diagonal quadrupole-quadrupole interaction. The results obtained in a space
that spans less than 10\% of the full-space reproduce the correct binding
energy (within 2\% of the full-space result) as well as the low-energy spectrum
and structure of the states that have greater than 90\% overlap with the
full-space results. In contrast, for a $m$-scheme spherical shell-model
calculation one needs about 60\% of the full space to obtain results comparable
with the oblique basis results.

Studies of the lower pf-shell nuclei, $^{44-48}Ti$ and $^{48}Cr$, using the
realistic Kuo-Brown-3 (KB3) interaction show strong SU(3) symmetry breaking due
mainly to the single-particle spin-orbit splitting. Thus the KB3 Hamiltonian
could also be considered a two-mode system. This has been further supported by
the behavior of the yrast band B(E2) values that seems to be insensitive to
fragmentation of the SU(3) symmetry. Specifically, the quadrupole collectivity
as measured by the B(E2) strengths remains high even though the SU(3) symmetry
is rather badly broken. This has been attributed to a quasi-SU(3) symmetry
where the observables behave like a pure SU(3) symmetry while the true
eigenvectors exhibit a strong coherent structure with respect to each of the
two bases. This provides the opportunity for further study of the implications
of two-mode calculations.

Future research may extend to multi-mode oblique calculations. An immediate
extension of the current scheme might use the eigenvectors of the pairing
interaction within the Sp(4) algebraic approach to the nuclear structure,
together with the collective SU(3) states and spherical shell model states.
Hamiltonian driven basis sets can also be considered. In particular, the method
may use eigenstates of the very-near closed shell nuclei obtained from a full
shell model calculation to form Hamiltonian driven J-pair states for mid-shell
nuclei. This type of extension would mimic the Interacting Boson Model (IBM)
and the so-called broken-pair theory. In particular, the three exact limits of
the IBM can be considered to comprise a three-mode system.  Nonetheless, the
real benefit of this approach is expected when the spaces encountered are too
large to allow for exact calculations.

\vspace{0.5cm}
{\it Acknowledgments.}
We acknowledge support from the U.S. National
Science Foundation under Grant No. PHY-9970769 and Cooperative Agreement No.
EPS-9720652 that includes matching from the Louisiana Board of Regents Support
Fund. V. G. Gueorguiev is grateful to the Louisiana State University Graduate
School for awarding him a dissertation fellowship and to the U. S. National
Science Foundation for the support for young scientists to attend the
International Nuclear Structure Conference on ``Mapping the Triangle'' held May
22-25, 2002 in Grand Teton National Park, Wyoming, and in so doing allowing him
to present the main results of his Ph.D. dissertation.

\end{document}